\documentclass{appolb}
\usepackage{epsfig}

\begin{document}

\title{Present status of nuclear cluster physics and
experimental perspectives
\thanks{Invited introductory Talk presented at the 3rd International Workshop on
"State of the Art in Nuclear Cluster Physics", May 26-30
2014 at Yokohama, Japan}}

\author{C. Beck$^a$
\address{
$^a$D\'epartement de Recherches Subatomiques, Institut Pluridisciplinaire 
Hubert Curien, IN$_{2}$P$_{3}$-CNRS and Universit\'e de Strasbourg - 23, rue 
du Loess BP 28, F-67037 Strasbourg Cedex 2, France\\
E-mail: christian.beck@iphc.cnrs.fr\\}
}

\maketitle

\newpage




\begin{abstract}

Knowledge on nuclear cluster physics has increased
considerably as nuclear 
clustering remains one of the most fruitful domains of nuclear physics, 
facing some of the greatest challenges and opportunities in the years ahead. 
The occurrence of ``exotic'' shapes in light $N$=$Z$ $\alpha$-like 
nuclei and the evolution of clustering from stability to
the drip-lines are being investigated more and more
accurately both theoretically and experimentally. 
Experimental progresses in understanding these questions
were recently examined and will be further revisited in
this introductory talk: clustering aspects are, in particular, discussed for 
light exotic nuclei with a large neutron excess such as neutron-rich Oxygen
isotopes with their complete spectrocopy.

\end{abstract}


\section{Introduction}

One of the greatest challenges in nuclear science is understanding the
structure of light nuclei from both experimental and theoretical perspectives.
Figure 1 was used by Catford to summarize the different types of clustering 
discussed during the last Cluster Conference in Debrecen~\cite{Catford12}.
Most of these structures were investigated in an experimental
context by using either some new approaches or 
developments of older methods ~\cite{Papka12}.

\begin{figure}
\includegraphics[scale= 0.65]{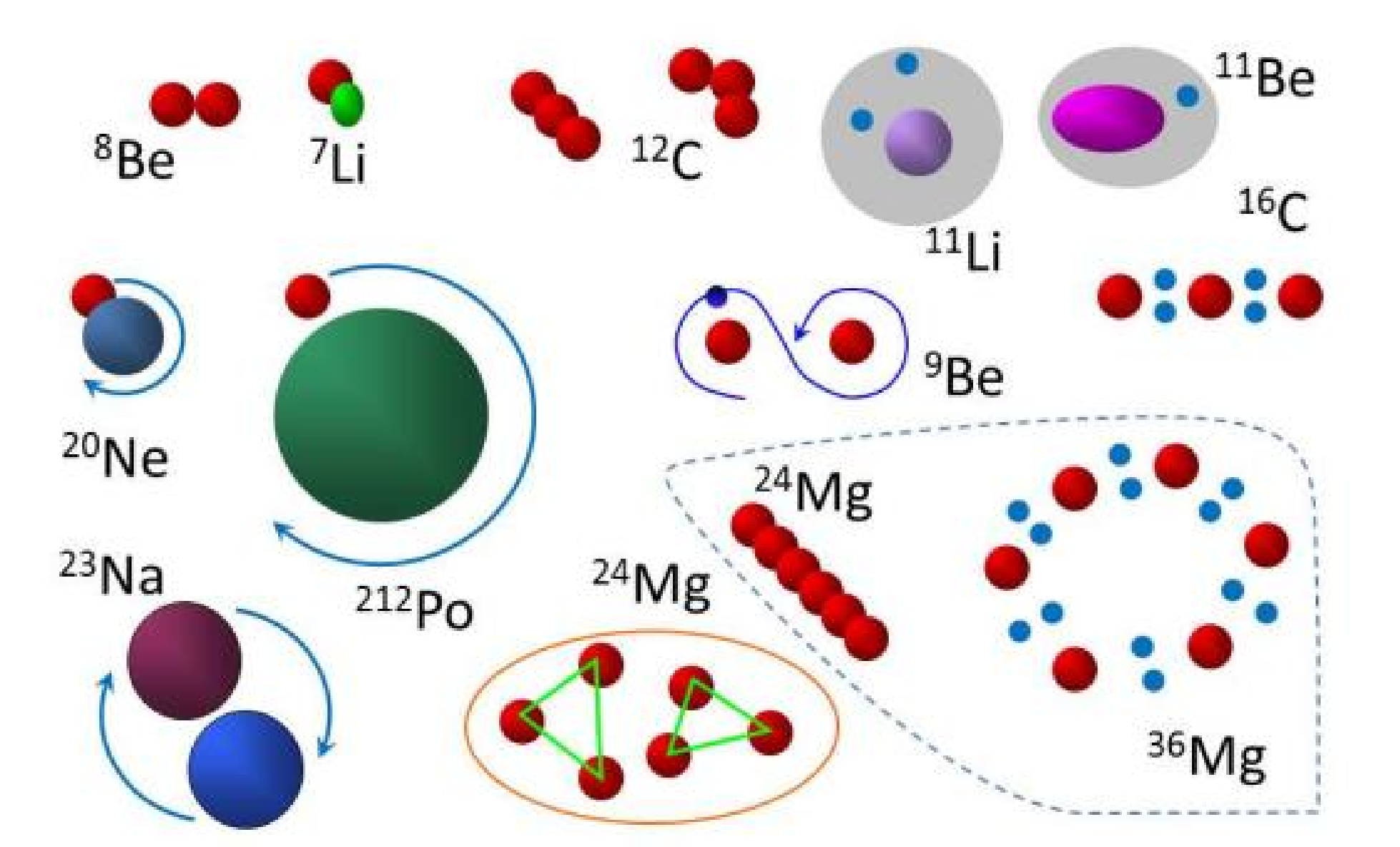}
\caption{\label{label} Different types of clustering in
nuclei that have been discussed at this workshop. 
(Figure adapted from Ref.\cite{Catford12} courtesy
from Catford).}
\label{fig:1}
\end{figure}

Starting in the 1960s the search for resonant structures in the excitation functions for various 
combinations of light $\alpha$-cluster ($N$=$Z$) nuclei in the energy regime 
from the Coulomb barrier up to regions with excitation energies of $E_{x}$=20$-$50~MeV 
remains a subject of contemporary debate~\cite{Greiner95,Beck94}. These 
resonances~\cite{Beck94} have been interpreted in terms of nuclear molecules~\cite{Greiner95}. 
The question of how quasimolecular resonances may reflect continuous transitions
from scattering states in the ion-ion potential to true cluster states in the 
compound systems was still unresolved in the 1990s~\cite{Greiner95,Beck94}. In many 
cases, these resonant structures have been associated with strongly-deformed 
shapes and with $\alpha$-clustering phenomena \cite{Freer07,Horiuchi10}, predicted from the 
Nilsson-Strutinsky approach, the cranked $\alpha$-cluster model~\cite{Freer07}, or 
other mean-field calculations~\cite{Horiuchi10,Gupta10}. In light $\alpha$-like 
nuclei clustering is observed as a general phenomenon at high excitation energy 
close to the $\alpha$-decay thresholds \cite{Freer07,Oertzen06}. This exotic 
behavior has been perfectly illustrated by the famous
''Ikeda-Diagram" for $N$=$Z$ 
nuclei in 1968 \cite{Ikeda}, which has been modified and extended by von Oertzen 
\cite{Oertzen01} for neutron-rich nuclei more than 10
years ago, as shown in the
left panel of figure 2.
Clustering is a general feature \cite{Milin14} not only
observed in typical light
neutron-rich nuclei \cite{Kanada10}, but also in less
common systems such as the neutron-halo $^{11}$Li 
\cite{Ikeda10} and/or $^{14}$Be nuclei, for instance \cite{Nakamura12}. The problem of 
cluster formation has also been treated extensively for very heavy systems by 
Gupta \cite{Gupta10}, by Zagrebaev and Greiner \cite{Zagrebaev10} and
by Simenel \cite{Simenel14} where giant molecules and collinear ternary 
fission may co-exist \cite{Kamanin14}. Finally, signatures of $\alpha$ clustering have
also been discovered in light nuclei undergoing ultrarelativistic nuclear 
collisions \cite{Zarubin14,Broniowski14}.
In this introductory talk, I will limit myself first to the light $^{12}$C
and $^{16}$O $\alpha$-like nuclei in Section 2, then to $\alpha$ clustering, deformations and 
$\alpha$ condensates in heavier $\alpha$-like nuclei in Section 3, and,
finally, to clustering in light
neutron-rich nuclei in Section 4.

\begin{figure}
\includegraphics[scale= 0.55]{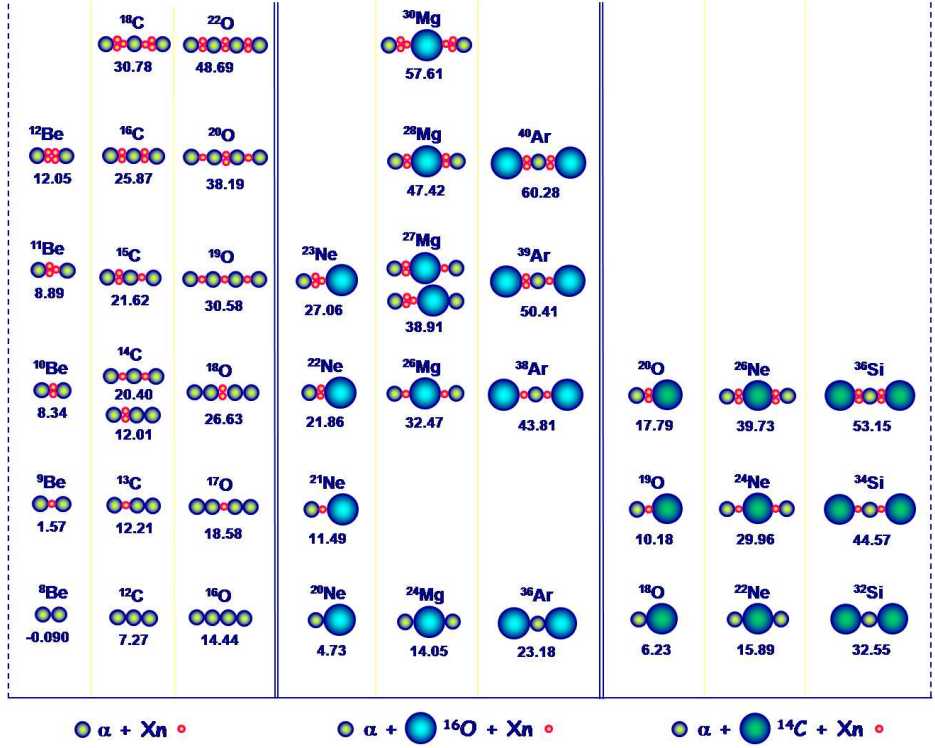}
\caption{\label{label}Schematic illustration of the structures of molecular
shape isomers in light neutron-rich isotopes of nuclei consisting
of $\alpha$-particles, $^{16}$O- and $^{14}$C-clusters plus some
covalently bound neutrons (Xn means X neutrons). The so called "Extended 
Ikeda-Diagram" \cite{Oertzen01} with $\alpha$-particles (left panel) and 
$^{16}$O-cores (middle panel) can be generalized to $^{14}$C-cluster cores 
(right panel). The lowest line of each configuration corresponds to parts
of the original "Ikeda-Diagram" \cite{Ikeda}. However, because of its deformation,
the $^{12}$C nucleus is not included, as it was earlier \cite{Ikeda}.
Decay threshold energies (in MeV) are given for the relevant 
decompositions of clusters. (Figure adapted from Ref.\cite{Milin14}
courtesy from von Oertzen).}
\label{fig:2}
\end{figure}

\section{Renewed interest in the spectroscopy of $^{12}$C
and $^{16}$O $\alpha$-like nuclei}
\label{sec:1}

The renewed interest in $^{12}$C was mainly focused to a better understanding 
of the nature of the so called "Hoyle" state \cite{Hoyle54} that can be described 
in terms of a bosonic condensate, a cluster state and/or a $\alpha$-particle 
gas \cite{Tohsaki01,Oertzen10a,Yamada}. Much experimental progress has been
achieved recently as far as the spectroscopy of  $^{12}$C
near and above the $\alpha$-decay threshold is
concerned~\cite{Itoh04,Freer11,Itoh11,Zimmerman13,Kokalova13a,Marin14}. More
particularly, the 2$^{+}_{2}$
"Hoyle" rotational excitation in  $^{12}$C has been observed
by several experimental groups \cite{Itoh04,Itoh11,Zimmerman13}.
The most convincing experimental result, displayed in
figure 3, comes from measurements of the $^{12}$C($\gamma$,$\alpha$)$^8$Be
reaction performed at the HIGS facility
\cite{Zimmerman13}. The angular distributions of the
alpha particles in the
region of 9-10 MeV are consistent with an L=2 pattern,
including a dominant 2$^+$ component. This 2$^{+}_{2}$
state that appears at around 10 MeV is considered to be the 2$^+$
excitation of the "Hoyle" state (in agreement with the
previous experimental investigation of Itoh et
al.~\cite{Itoh04}) according to the $\alpha$
cluster \cite{Uegaki} and $\alpha$ condensation models
\cite{Tohsaki01,Yamada}. 
On the other hand, the experiment
$^{12}$C($\alpha$,$\alpha$)$^{12}$C$^*$ carried out at the
Birmingham cyclotron 
~\cite{Marin14}, UK, populates a new state compatible with an
equilateral triangle configuration of three $\alpha$ particles.
Still, the structure of the "Hoyle" state remains controversial
as experimental results of its direct decay into three
$\alpha$ particles are found to be in disagreement
~\cite{Freer94,Raduta11,Manfredi12,Kirsebom12,Rana13,Itoh14}.

\begin{figure}
\includegraphics[scale=0.75]{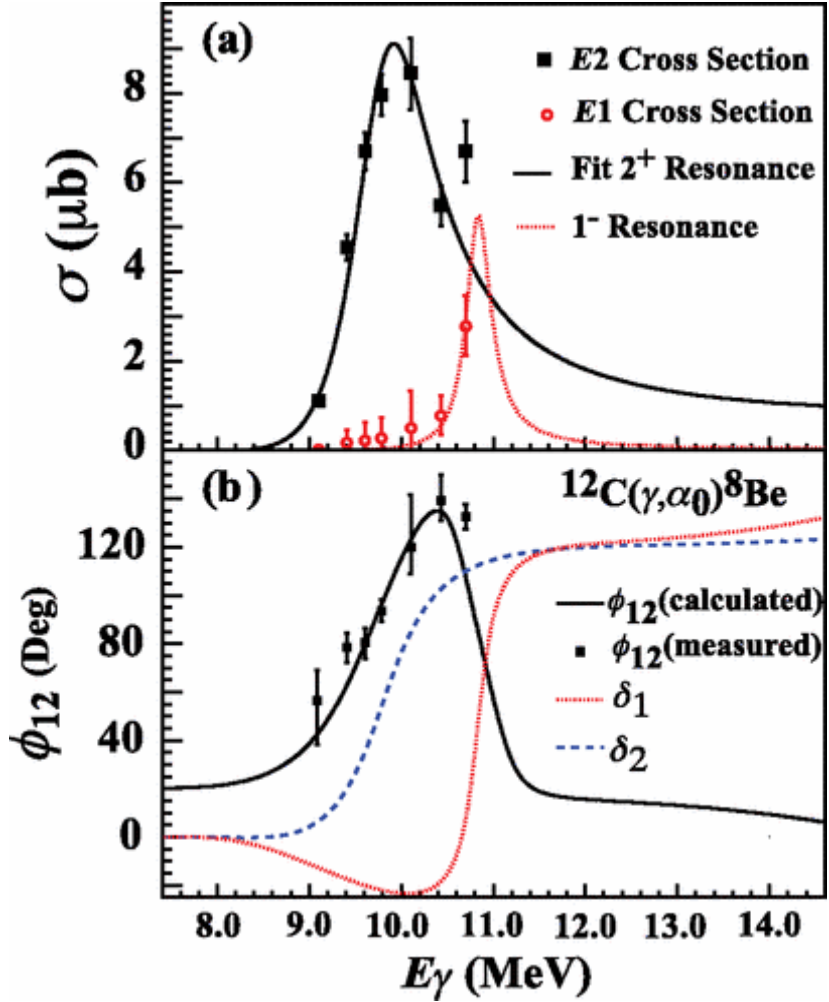}
\caption{\label{label}a) Measured E1 and E2 cross sections of
$^{12}$C($\gamma$,$\alpha_0$)$^8$Be. b)
Measured E1 and E2 phase angles. (Figure taken from Ref.
\cite{Zimmerman13} courtesy from Gai).}
\label{fig:3}
\end{figure}

\begin{figure}
\includegraphics[scale=0.60]{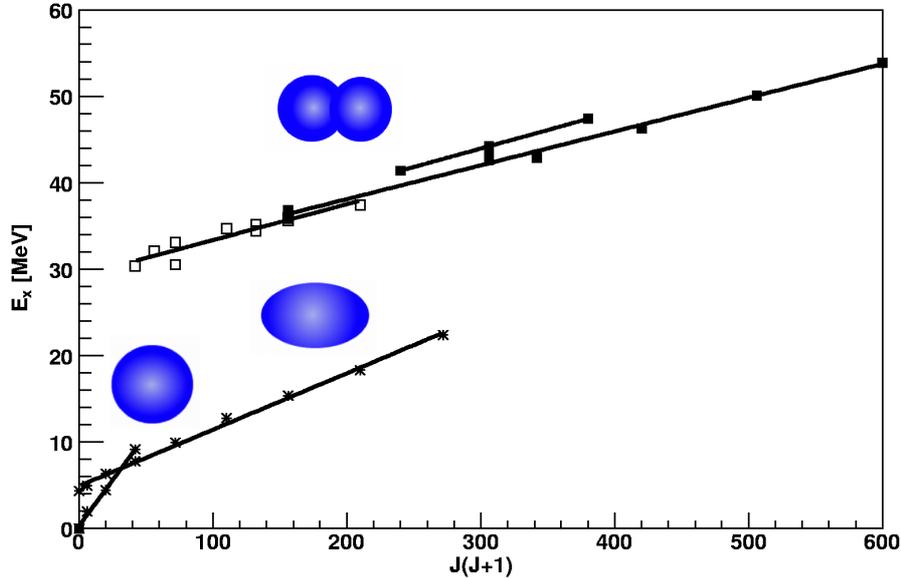}
\caption{\label{label}Experimental rotational bands and
schematical deformed shapes in $^{36}$Ar. Excitation energies 
	of the ground state (spherical shape) and SD (ellipsoidal shape) bands~\cite{Svensson00}, respectively, 
	and the energies of HD (dinuclear shape) band from 
	the quasimolecular resonances observed in the $^{12}$C+$^{24}$Mg 
	(open rectangles)  and  $^{16}$O+$^{20}$Ne (full rectangles) reactions 
	are plotted as a function of J(J+1). The relevant
	energies of the resonance can be found in
	Refs.~\cite{Beck11,Beck13,Beck14}. (Figure
	reproduced from Refs.~\cite{Beck11}).}
\label{fig:4} 
\end{figure}

In the study of Bose-Einstein Condensation (BEC) the
$\alpha$-particle states in light $N$=$Z$ nuclei \cite{Tohsaki01,Oertzen10a,Yamada}, 
are of great importance. At present, the search for an experimental signature of 
BEC in $^{16}$O is of highest priority. 
A state with the structure of the ''Hoyle" state \cite{Hoyle54} in 
$^{12}$C coupled to an $\alpha$ particle is 
predicted in  $^{16}$O at about 15.1 MeV (the 0$^{+}_{6}$ state), the
energy of which is $\approx$ 700 keV above the 4$\alpha$-particle 
breakup threshold \cite{Funaki08}. However, any state in
$^{16}$O equivalent 
to the ''Hoyle" state \cite{Hoyle54} in $^{12}$C is most certainly 
going to decay exclusively by particle emission with very small 
$\gamma$-decay branches, thus, very efficient
particle-$\gamma$ coincidence techniques will 
have to be used in the near future to search for them. BEC states are expected to 
decay by alpha emission to the ''Hoyle" state and could be found among the
resonances in $\alpha$-particle inelastic scattering on $^{12}$C decaying to 
that state. In 1967 Chevallier et al.
\cite{Chevallier67,Brochard76} could excite these states in an $\alpha$-particle 
transfer channel leading to the 
$^{8}$Be--$^{8}$Be final state and proposed that a
structure corresponding to a rigidly rotating linear
arrangement of four alpha particles may exist in $^{16}$O.
Very recently, a more sophisticated experimental setup was used at Notre
Dame \cite{Curtis13}: although the excitation function is
generally in good agreement with the previous results \cite{Chevallier67,Brochard76}
a phase shift analysis of the angular distributions does
not provide evidence to support the reported
hypothesis of a 4$\alpha$-chain state configuration.
Experimental investigations are still underway to
understand the nuclear structure of high spin states of
both $^{16}$O and
$^{20}$Ne nuclei for instance at Notre Dame
\cite{Kokalova13b} and/or
iThemba Labs \cite{Papka14} facilities.
Another possibility might be to perform Coulomb 
excitation measurements with intense $^{16}$O and
$^{20}$Ne beams at intermediate energies.

\section{Alpha clustering, deformations and alpha
condensates in heavier nuclei}
\label{sec:1}

The real link between superdeformation (SD), nuclear molecules and alpha 
clustering \cite{Horiuchi10,Beck04a,Beck04b,Cseh09,Beck11} is of 
particular interest, since nuclear shapes with major-to-minor axis ratios of 
2:1 are typical ellipsoidal elongations for light nuclei
(corresponding to a quadrupole deformation 
parameter $\beta_2$ $\approx$ 0.6). Furthermore, the structure 
of possible octupole-unstable 3:1 nuclear shapes (hyperdeformation (HD) with $\beta_2$ $\approx$ 
1.0) has also been 
discussed for actinide nuclei \cite{Cseh09} in terms of clustering phenomena. Typical examples for 
the possible relationship between quasimolecular bands and extremely deformed (SD/HD) 
shapes have been widely discussed in the literature for $A = 20-60$ $\alpha$-conjugate 
$N$=$Z$ nuclei, such as $^{28}$Si 
\cite{Taniguchi09,Ichikawa11,Jenkins12,Jenkins13,Darai12}, $^{32}$S 
\cite{Horiuchi10,Ichikawa11,Kimura04,Lonnroth10,Chandana10}, 
$^{36}$Ar
\cite{Cseh09,Beck11,Svensson00,Beck08a,Beck09,Beck13,Beck14}, $^{40}$Ca 
\cite{Ideguchi01,Rousseau02,Taniguchi07,Norrby10}, $^{44}$Ti 
\cite{Horiuchi10,Leary00,Fukada09}, $^{48}$Cr \cite{Salsac08} and $^{56}$Ni 
\cite{Nouicer99,Rudolph99,Beck01,Bhattacharya02,Oertzen08,Wheldon08}.

Highly deformed shapes and SD rotational bands have been 
discovered in several light $\alpha$-conjugate ($N$=$Z$) nuclei, such as $^{36}$Ar
and $^{40}$Ca by using $\gamma$-ray spectroscopy techniques 
\cite{Svensson00,Ideguchi01}. In particular, the extremely deformed rotational
bands in $^{36}$Ar \cite{Svensson00} (shown as crosses in
figure~4) might be 
comparable in shape to the quasimolecular bands observed in both $^{12}$C+$^{24}$Mg 
(shown as open rectangles in figure~4)
and $^{16}$O+$^{20}$Ne (shown as full
rectangles in figure~4) reactions. Ternary clusterizations are also predicted 
theoretically, but were not found experimentally in $^{36}$Ar so far 
\cite{Beck09}. On the other hand, ternary fission of $^{56}$Ni -- related to its hyperdeformed 
(HD) shapes -- was claimed to be identified from out-of-plane angular correlations 
measured in the $^{32}$S+$^{24}$Mg reaction with the Binary Reaction 
Spectrometer (BRS) at the {\sc Vivitron} Tandem facility of the IPHC, Strasbourg 
\cite{Oertzen08}, though this remains to be confirmed
\cite{Wheldon08}. This possibility \cite{Oertzen08} is not limited to light 
$N$=$Z$ compound nuclei, true ternary fission
\cite{Zagrebaev10,Kamanin14,Pyatkov10}
can also occur for very heavy \cite{Kamanin14,Pyatkov10} and superheavy 
\cite{Zagrebaev10b} nuclei.

The next natural question to be addressed is whether
dilute-gas-like structures (i.e. BEC)
\cite{Tohsaki01,Oertzen10a,Yamada}
also exist in medium-mass
$\alpha$-conjugate nuclei as predicted by several
theoretical investigations
\cite{Yamada04,Khan,Girod}. Several recent undergoing
experiments indicate that it might be the case at least
for $^{24}$Mg
\cite{Kawabata13,Francesca14}, $^{36}$Ar \cite{Akimone14}
and
$^{56}$Ni \cite{Akimone13} and much work is in progress
in this field \cite{Oertzen10a}.

\begin{figure}
\includegraphics[scale=0.62]{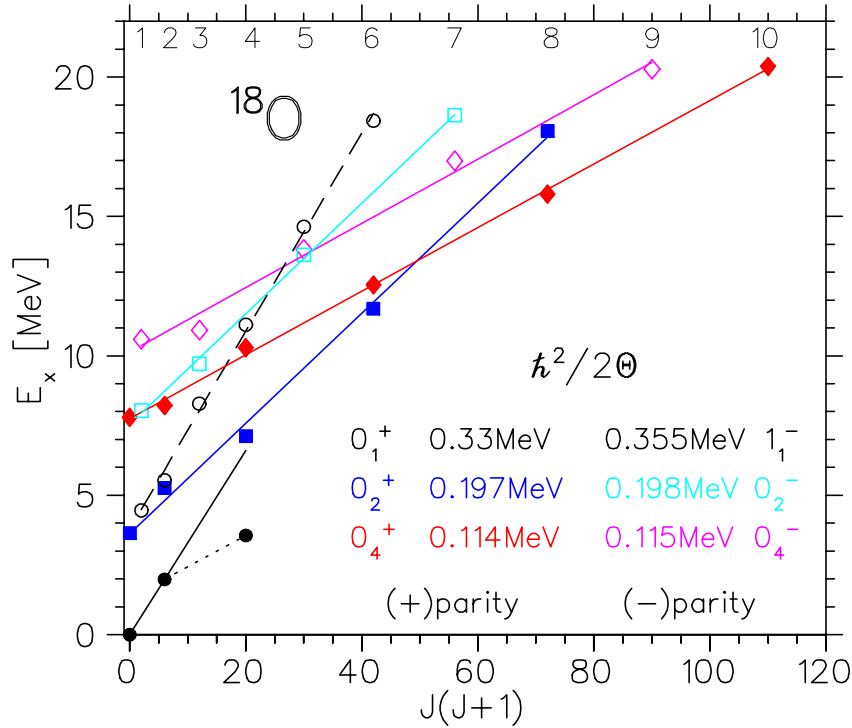}
\caption{\label{label}Overview of six rotational band
structures observed in $^{18}$O.
	Excitation energy systematics for the members of the rotational
	bands forming inversion doublets with K=0 are plotted as a function 
	of J(J+1). The curves are drawn to guide the eye for the slopes.
	The indicated slope parameters contain information on
	the moments of inertia. Square and open circle symbols correspond to 
	cluster bands, whereas diamonds symbols correspond to molecular 
	bands.(Figure adapted from \cite{Oertzen10b} 
	courtesy from von Oertzen).}
\label{fig:5}
\end{figure}

\begin{figure}
\includegraphics[scale=0.62]{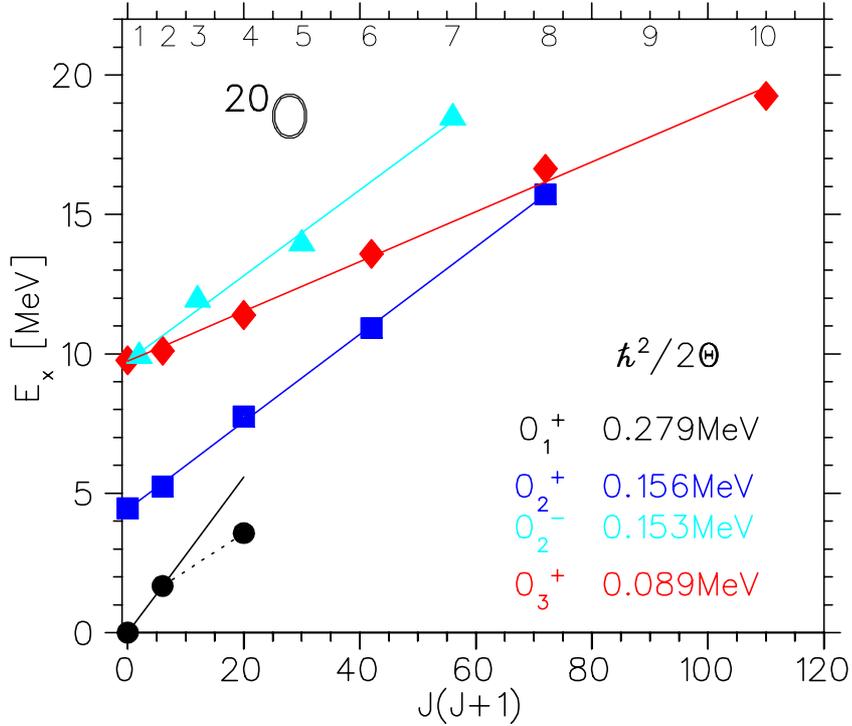}
\caption{\label{label}Overview of four rotational band structures observed in $^{20}$O.
	Excitation energy systematics for the members of the rotational
	bands forming inversion doublets with K=0 are plotted as a function 
	of J(J+1). The curves are drawn to guide the eye for the slopes.
	The indicated slope parameters contain information on
	the moments of inertia. Square and triangle symbols correspond to 
	cluster bands, whereas diamonds symbols correspond to molecular 
	bands.(Figure adapted from \cite{Oertzen09}
	courtesy from von Oertzen).}
\label{fig:6} 
\end{figure}

\section{Clustering in light neutron-rich nuclei}
\label{sec:2}

As discussed previously, clustering is a general phenomenon observed also in 
nuclei with extra neutrons as it is presented in an extended "Ikeda-Diagram" \cite{Ikeda} 
proposed by von Oertzen \cite{Oertzen01} (see the left and
middle panels of figure~2).
With additional neutrons, specific molecular structures appear with binding effects based 
on covalent molecular neutron orbitals. In these diagrams $\alpha$-clusters and 
$^{16}$O-clusters are the main ingredients. Actually, the $^{14}$C nucleus may 
play a similar role in clusterisation as the $^{16}$O
nucleus does. Both of them have similar properties 
as a cluster: i) closed neutron p-shells, ii) first excited states are well above 
E$^{*}$ = 6 MeV, and iii) high binding energies for $\alpha$-particles.

A general picture of clustering and molecular configurations in light nuclei 
can be drawn from a detailed investigation of the oxygen isotopes with
A $\geq$ 17. Here I will only present recent results on the even-even 
oxygen isotopes: $^{18}$O \cite{Oertzen10b} and $^{20}$O \cite{Oertzen09}. 
But very striking cluster states have also been found in odd-even oxygen 
isotopes such as: $^{17}$O \cite{Milin09} and $^{19}$O \cite{Oertzen11}. 

Figure 5 gives an overview of all bands in $^{18}$O as a plot of excitation energies
as a function of J(J+1) together with their respective moments of inertia. In the
assignment of the bands both the dependence of excitation energies on J(J+1)
and the dependence of measured cross sections on 2J+1 \cite{Oertzen10b}
were considered. Slope parameters obtained in
a linear fit to the excitation energies \cite{Oertzen10b} indicate the moment
of inertia of the rotational bands given in figure~5. The intrinsic structure
of the cluster bands is reflection asymmetric, the parity projection gives an 
energy splitting between the partner bands.  
The assignment of the experimental molecular bands are supported by either
generator-coordinate-method \cite{Descouvemont} or Antisymmetrized Molecular
Dynamics (AMD) calculations \cite{Furutachi08}. 

The bands of $^{20}$O \cite{Oertzen09}
shown in figure~6 can be compared with those of $^{18}$O
displayed in figure.~5. The first doublet (K=0$^{\pm}_{2}$) has a slightly larger moment of
inertia (smaller slope parameter) in $^{20}$O, which is consistent
with its interpretation as $^{14}$C--$^{6}$He or $^{16}$C--$^{4}$He 
molecular structures (they start well below the thresholds of 16.8 MeV and 
12.32 MeV, respectively). The second band, for which the negative parity 
partner has yet to be determined, has a slope parameter slightly smaller
than in $^{18}$O. This is consistent with the study of the bands in 
$^{20}$O by Furutachi et al. \cite{Furutachi08}, which clearly establishes 
parity inversion doublets predicted by AMD calculations for the 
$^{14}$C--$^6$He quasimolecular (cluster) and $^{14}$C-2n-$\alpha$ molecular structures.
The corresponding moments of inertia illustrated in figure~4 and
figure ~5 are 
strongly suggesting large deformations for the cluster
structures. It may be
concluded that the  reduction of the moments of inertia of the lowest
bands of $^{18,20}$O is consistent with the assumption that the strongly bound $^{14}$C 
nucleus have equivalent properties to $^{16}$O. It is
interesting to note that the Quantum Mechanical
Fragmentation Theory (QMFT) of Gupta \cite{Gupta10} reaches
to the same conclusion on the possibility of $^{14}$C
clustering \cite{Bansai11}.
Therefore, the "Ikeda-Diagram" 
\cite{Ikeda} and the "Extended Ikeda-Diagram" consisting of $^{16}$O cluster
cores with covalently bound neutron \cite{Oertzen01} must be further extended to 
include also the $^{14}$C cluster cores as illustrated in
right panel of
figure 2.

\section{Summary and outlook}

The connection of $\alpha$-clustering, quasimolecular resonances, $\alpha$ condensates 
in very light nuclei and extreme deformations (SD, HD, ...) in heavier nuclei as investigated 
by more and more sophisticated experimental devices, has been discussed in this 
introductory talk. In particular, high-precision spectroscopy techniques
unable us to uncover important parts of the complete spectroscopy of the so called 
"Hoyle" state in $^{12}$C. The origin of carbon for life is likely to be 
understood in the very near future with answer to the question of the "Hoyle" state structure.
Similarly the quest for the  4$\alpha$ states of $^{16}$O near the $^{8}$Be+$^{8}$Be 
and $^{12}$C+$\alpha$ decay thresholds, which correspond to the ''Hoyle" state. 
Results have also been presented on neutron-rich oxygen isotopes displaying very well defined 
quasimolecular bands in agreement with AMD predictions. Consequently, the 
"Extended Ikeda-Diagram" has been further extended for light neutron-rich nuclei by 
inclusion of the $^{14}$C cluster as a core, similarly to the $^{16}$O one. 
The search for extremely elongated configurations (HD) in rapidly rotating medium-mass 
nuclei, which has been pursued by $\gamma$-ray spectroscopy measurements, will have 
to be performed in conjunction with charged-particle techniques in the near future 
since such states are most likely going to decay by
particle emission (see \cite{Papka12,Oertzen08,Wheldon08}).

\section{Acknowledments}

I would like to thank the Sotancp3 organizers for giving me the opportunity
to introduce this workhop. Christian Caron (Springer) is
acknowledged for 
initiating the new series of three volumes of \emph{Lecture Notes in Physics} dedicated 
to "Clusters in Nuclei".


\end{document}